# Challenges and open problems in computational prediction of protein complexes: the case of membrane complexes

Sriganesh Srihari

Department of Computer Science, National University of Singapore, Singapore 117590.

Present affiliation: Institute for Molecular Bioscience, The University of Queensland, St. Lucia, Queensland 4067, Australia.

**Abstract**

Identifying the entire set of complexes is essential not only to understand complex formations, but also to map the high level organisation of the cell. Computational prediction of protein complexes faces several challenges including the lack of sufficient protein interactions, presence of noise in protein interaction datasets and difficulty in predicting small and sparse complexes. These challenges are covered in most reviews of complex prediction methods. However, an important challenge that needs to be addressed is the prediction of membrane complexes. These are often ignored because existing protein interaction detection techniques do not detect interactions between membrane proteins. But, recently there have been several new experimental techniques including MY2H that are capable of detecting membrane protein interactions. In the light of this new data, we discuss here new challenges and the kind of open problems that need to be solved to effectively detect membrane complexes.

## 1. Background

Most biological processes are carried out by proteins that physically interact to form multiprotein complexes. These complexes are comprised of many subunits that work in a coherent fashion, and interact with individual proteins or other complexes to form functional modules and pathways that drive the cellular machinery. Therefore, identifying the entire set of complexes (the "complexosome") is essential to not only understand complex formations, but also to map the high level organization of the cell (for reviews, see Srihari et al., 2010 – 2013).

Among these, *membrane protein complexes* are formed by physical interaction among membrane proteins. Membrane proteins are attached to or associated with the membranes of the cell or its organelles. Membrane proteins constitute approximately 30% of the proteomes of organisms. Begin hydrophobic, membrane proteins are the least studied subset of proteins because these are difficult to crystallize using traditional X-ray crystallography compared to their soluble counterparts.

Membrane proteins are involved in the transportation of ions, metabolites and larger molecules such as proteins, RNA and lipids across membranes, in sending and receiving of chemical signals, in propagating electrical impulses, in anchoring enzymes and other proteins, in controlling membrane lipid composition, and in organizing and maintaining the shape of organelles and the cell itself (Heijne et al., 2007). In humans, membrane proteins involved in signal transduction across membranes, the so-called G-protein-coupled-receptors (GPCRs), alone account for 15% of the membrane proteins, and *30% of all drug targets* are GPCRs (Heijne et al., 2007). Due to the key roles of membrane proteins, the identification of their interactions and understanding their complex formations is of particular interest in understanding diseases and in aiding new drug development.

Membrane protein complexes are notoriously difficult to study using traditional high-throughput techniques (Lalonde et al., 2008). Intact membrane complexes are difficult to "pull down" using conventional affinity purification/mass spectrometry-based systems. This is due in part to the hydrophobic nature of membrane proteins, as well as the ready dissociation of subunit interactions, either between transmembrane subunits or between transmembrane and cytoplasmic subunits (Barrera et al., 2008). Further, membrane protein structure is difficult to study by traditional high-resolution methods like X-ray crystallography and NMR spectroscopy.

One major avenue is to understand membrane proteins and their complexes is to map the membrane protein 'subinteractomes', the set of interactions among membrane proteins. Conventional Yeast two-hybrid (Y2H) system used to detect binary interactions utilizes the observation that the DNA-binding domain and the activation domain of a transcription factor can associate and activate transcription despite their fusion to different proteins as long as they are in proximity (Ito et al., 2001; Uetz et al., 2000). This transcriptional activation is observed through the use of an appropriate reporter gene. Although this system is both powerful and robust, the interaction is confined to the nucleus of the cell thereby excluding the study of membrane proteins.

In order to counter the disadvantages of conventional techniques, new biochemical techniques have been developed recently to facilitate the characterization of interactions among membrane proteins. Among these is the split-ubiquitin membrane yeast two-hybrid (MYTH) system (Miller et al., 2005; Kittanakom et al., 2009;

Staglijar et al., 2010; Petschnigg et al., 2011). This system is based on ubiquitin, an evolutionarily conserved 76 amino acid protein that serves as a tag for proteins targeted for degradation by the 26S proteasome. The presence of ubiquitin is recognized by ubiquitin specific proteases (UBPs) located in the nucleus and cytoplasm of all eukaryotic cells. Ubiquitin can be split and expressed as two halves, the amino-terminal and the carboxyl terminal. These two halves have a high affinity for each other in the cell and can reconstitute to form pseudo-ubiquitin that is recognizable by UBPs.

In MYTH, the C-terminal is fused to a protein of interest ('bait') and an artificial transcription factor (TF), and the 'prey' proteins are fused to the N-terminal. The two halves reconstitute into a pseudo-ubiquitin protein if there is affinity between the bait and prey proteins. This pseudo-ubiquitin is recognized by UBPs, which cleaves after the C-terminus of ubiquitin to release the TF, which then enters the nucleus to activate the reporter genes.

Large-scale screens are performed by transforming a 'bait'-containing yeast strain with a library of 'prey' plasmids and plating transformants on media that selects for cells that have activated the reporter genes of the system. Interactions are confirmed through a series of assays including 'bait'-dependency tests and co-immunoprecipitation (Kittanakom et al., 2009).

With the development of this split ubiquitin system, a fair number of interactions among membrane proteins are being catalogued: 343 interactions among 179 proteins (Lalonde et al., 2010), 808 interactions among 536 proteins (Miller et al., 2005). The need now is to develop sophisticated algorithms to mine membrane complexes from these interactions.

The need now is to develop effective computational techniques to mine this membrane subinteractome data to reconstruct membrane complexes. These computation techniques can become vital alternatives for experimental techniques in identifying membrane complexes, and thereby aiding in completing the "complexosome" map, and also for studying potential drug targets.

## 2. Challenges in identifying membrane complexes using traditional/existing methods

The identification of membrane complexes requires understanding their assembly – how the individual proteins come together to form complexes, and how these complexes are eventually degraded. This is because membrane proteins are not stable entities as their soluble counterparts.

Studies reveal that this assembly occurs in an orderly fashion, that is, membrane complexes are formed by an ordered assembly of intermediaries, and in order to prevent unwanted intermediaries, this assembly is highly aided by chaperones (Daley et al., 2008). Why membrane complexes assemble in an ordered manner is unclear, but studies suggest that this could be a protection mechanism of the cell against harmful intermediaries (Harrmann et al., 2005).

Many membrane complexes are formed by transient interactions involving exchange of proteins in and out of existing complexes – a phenomenon called 'dynamic exchange' (Daley et al., 2008). Such exchange has been shown for the Translocase of the mitochondrial Outer Membrane (TOM complex) and the NADH-ubiquinone oxidoreductase (Model et al., 2001; Lazarou et al., 2007). Dynamic exchange is thought to be a mechanism for regulating proteins within membrane complexes.

In order to identify and analyse membrane complexes in-depth it is important to take into account the above findings. This will involve identification of membrane complexes from the yeast membrane subinteractome, followed by incorporation of chaperone and, if available, 'time' information to study the assembly of membrane complexes. This research proposal is about developing computational methods to perform such analyses.

### 3.     Open computational problems

a) How to extract out the membrane 'subinteractome'?
- Very few membrane complexes characterised for PPIs.

b) How to filter these interactions (since the network will already be sparse)?
- Combine it with the whole network or neighbors among soluble proteins for this process?

c) Do clusters correspond to membrane complexes?
- Obstacle: Too many transient interactions.
- There is dynamic exchange of proteins happens across complexes. So, different "isophorms" of the same complex should be detected?
- Where to obtain "time" information for this and how to incorporate into the clustering?

d) How to evaluate the complexes, since very few *bona fide* complexes are known?

- All the more interesting because we can provide putative complexes for validation!
- We can also assign roles to unannotated proteins.

e) Studying the relationships (complex formations) between membrane proteins and soluble proteins.

# References

1. Barrera, N.P., Bartolo, N.D., Booth, P.J., Robinson, C.V.: **Micelles Protect Membrane Complexes from Solution to Vacuum.** *Science* 2008, **321**: 243.

2. Gavin, A.C., Aloy, P., Grandi, P., Krause, R., et al.: **Proteome survey reveals modularity of the yeast cell machinery.** *Nature* 2006, **440**: 631-636.

3. Gavin, A.C., Bosche, M., Krause, R., Grandi, P., Marzioch, M., et al.: **Functional organization of the yeast proteome by systematic analysis of protein complexes.** *Nature* 2002, **415**: 141-147.

4. Ito, T., Chiba, T., Ozawa, R., Yoshida, M., Hattori, M., Sakaki, Y., et al.:
**A comprehensive two-hybrid analysis to explore the yeast protein interactome.** *Proc. Natl. Acad. Sci.* 2001, **98**: 4569-4574.

5. Kittanakom, S., Chuk, M., Wong, V., Snyder, J., Edmonds, D., Lydakis, A., Zhang, Z., Auerbach, D., Stagljar, I.: **Analysis of membrane protein complexes using the split-ubiquitin membrane yeast two-hybrid (MYTH) system.** *Methods Mol. Bio.* 2009, **548**: 247.

6. Krogan, N.J., Cagney, G., Yu, H., Zhong, G., et al.: **Global landscape of protein complexes in the yeast Saccharomyces cerevisiae.** *Nature* 2006, **440**: 637-643.

7. Lalonde, S. Ehrhardt, D.W., Logue, D., Chen, J., Rhee, S.Y., Frommer, W.B.: **Molecular and cellular approaches for the detection of protein-protein interactions: latest techniques and current limitations.** *Plant J.* 2008, **4**: 610-635.

8. Li, X.L., Wu, M., Kwoh, C.C., Ng, S-K.: **Computational approaches for detecting protein complexes from protein interaction networks: a survey.** *BMC Genomics* 2010, **11(S3).**

9. Miller, J.P., Russel.S. Ben-Hur, A., Desmarais, C., Stagljar, I., Novel, W.S., Fields, S.: **Large-scale identification of yeast integral membrane protein interactions.** *Proc Natl Acad Sci* 2005, **102 (34)**: 12123-12128.

10. Petshnigg, J., Moe, O.W., Stagljar, I.: **Using yeast as a model to study membrane proteins.** *Current Opinion in Nephrology and Hypertension* 2011, **20**: 425-432.

11. Shoemaker, B.A., Panchenko, A.R.: **Deciphering protein-protein interactions.** *PLoS Comp Bio* 2007, **3.**

12. Snider, J., Kittanakom, S., Damjanovic, D., Curak, J., Wong, V., Stagljar, I.: **Detecting interactions with membrane proteins using a membrane two-hybrid assay in yeast.** *Nature Protocols* 2010, **5(7)**: 1281.

13. Uetz, P., Giot, L., Cagney, G., Mansfield, T.A., Judson, R.S., Knight, J.R., et al.: **A comprehensive analysis of protein-protein interactions in Saccharomyces cerevisiae.** *Nature* 2000, **403**: 623-627.

14. von Heijne, G.: **The membrane protein universe: what's out there and why bother?** *J Internal Med.* 2007.

15. Srihari S, Ning K, Leong HW: MCL-CAw: a refinement of MCL for detecting yeast complexes from weighted PPI networks by incorporating core-attachment structure, *BMC Bioinformatics* 2010, 11(1):504.

16. Srihari S, Ning K, Leong HW: Refining Markov Clustering for protein complex prediction by incorporating core-attachment structure, *Genome Informatics*, World Scientific 2009.

17. Srihari S, Leong HW. Employing functional interactions for characterisation and detection of sparse complexes from yeast PPI networks. *Int J Bioinfo Res Appl* 2012, 8(3/4): 286-304.

18. Srihari S, Leong HW. A survey of computational methods for protein complex prediction from protein interaction networks. *J Bioinfo Comp Biol* 2013, 11(2): 1230002.

19. Srihari S, Leong HW. Temporal dynamics of protein complexes: a case study using yeast protein complexes. *BMC Bioinformatics* 2012, 13 (Suppl 17): S16.

20. Srihari S & Ragan MA. Systematic tracking of dysregulated modules identifies novel genes in cancer. *Bioinformatics* 2013, 29(12):1553-61.

21. Srihari S. Integrating biological insights with topological characteristics for improved complex prediction from protein interaction networks. PhD Thesis, National University of Singapore 2012.